\documentclass[12pt]{article}

\usepackage{epsfig}
\usepackage{amssymb,amsmath}

\newcommand{\bea}{\begin{eqnarray}}
\newcommand{\eea}{\end{eqnarray}}
\def\tr{\mathrm{tr}}
 \makeatletter
\def\alt{\mathrel{\mathpalette\gl@align<}}
\def\agt{\mathrel{\mathpalette\gl@align>}}
\def\gl@align#1#2{\lower.6ex\vbox{\baselineskip\z@skip\lineskip\z@
\ialign{$\m@th#1\hfil##\hfil$\crcr#2\crcr\sim\crcr}}} \makeatother

\begin{document}
\begin{flushright}
BA-08-01 \\
\end{flushright}
\vspace*{1.0cm}

\begin{center}
\baselineskip 20pt {\Large\bf
Higgs Mass Bounds, Type II SeeSaw and LHC
 } \vspace{1cm}

{\large Ilia Gogoladze$^{a,}$\footnote{ E-mail:
ilia@physics.udel.edu}, Nobuchika Okada$^{b,}$\footnote{ E-mail:
okadan@post.kek.jp} and Qaisar Shafi$^{a,}$\footnote{ E-mail:
shafi@bartol.udel.edu} } \vspace{.5cm}

{\baselineskip 20pt \it
$^a$Bartol Research Institute, Department of Physics and Astronomy, \\
University of Delaware, Newark, DE 19716, USA \\
\vspace{2mm} $^b$Theory Division, KEK, Tsukuba 305-0801, Japan }
\vspace{.5cm}

\vspace{1.5cm} {\bf Abstract}
\end{center}

In type II seesaw utilized to explain the observed neutrino
masses and mixings, one extends the Standard Model (SM) by
introducing scalar fields which transform as a triplet under the
electroweak gauge symmetry. New scalar couplings involving the Higgs
doublet then appear and, as we show, these have important
implications for the Higgs boson mass bounds obtained using vacuum
stability and perturbativity arguments. We identify, in particular,
regions of the parameter space which permit the SM Higgs boson to be
as light as 114.4 GeV, the LEP2 bound. The triplet scalars include
doubly charged particles whose masses could, in principle, be in the
few hundred GeV range, and so they may be accessible at the LHC.
\thispagestyle{empty}

\newpage

\addtocounter{page}{-1}

\baselineskip 18pt

The discovery of the Standard Model (SM) Higgs boson is arguably the
single most important mission for the LHC. Under the
somewhat radical assumption that the next energy frontier
lies at Planck scale ($M_{Pl}$), it has been found that the SM Higgs
boson mass lies in the range 127 GeV $\leq M_H \leq$ 170 GeV
\cite{stability1}. Here the lower bound of 127 GeV on $M_H$ derives
from arguments based on the stability of the SM vacuum (more
precisely, that the Higgs quartic coupling does not turn negative at
any scale between $M_Z$ and $M_{Pl}$). Thus, from the point of view
of the SM it is perhaps not too surprising that the Higgs boson has
not yet been found. The upper bound of about 170 GeV on
$M_H$ comes from the requirement that the Higgs quartic coupling
remains perturbative and does not exceed $\sqrt{4 \pi}$, say, during
its evolution between $M_Z$ and $M_{Pl}$.

It has become abundantly clear in recent years that an extension
 of the SM is needed to explain a number of experimental observations.
These include solar and atmospheric neutrino oscillations
\cite{NuData}, existence of non-baryonic dark matter \cite{WMAP},
the observed baryon asymmetry of the universe, etc.
Neutrino oscillations, in particular, cannot be understood within the SM,
even after including dimension five operators with Planck scale cutoff.
These operators yield neutrino masses of order $10^{-5}$ eV or less,
which is far below the 0.05-0.01 eV scale needed to explain
the observed atmospheric and solar neutrino oscillations, respectively.

Two attractive seesaw mechanisms exist for explaining the measured
neutrino masses (more accurately, mass differences squared). In the
so-called type I seesaw \cite{seesawI}, new physics is added to the
SM by introducing at least two heavy right-handed neutrinos. The
seesaw mechanism then ensures that the observed neutrinos acquire
masses which are suppressed by the heavy right-handed neutrino mass
scale(s). One expects that the heaviest right-handed neutrino has a
mass less than or of order $10^{14}$ GeV (This, roughly speaking,
comes from the seesaw formula $m_D^2/M_R$ for the light neutrino
mass, where $m_D$ and $M_R$ denote the Dirac and right-handed
neutrino masses, respectively, and assumes that $m_D$ is
less than or of order the electroweak scale).

In type II seesaw \cite{seesawII}, the SM is supplemented by a
SU(2)$_L$ triplet scalar field $\Delta$ which also carries
unit hypercharge. There exist renormalizable couplings
$\ell^T \Delta \ell$ which enable the neutrinos to acquire their
tiny (observed) masses through the non-zero VEV of $\Delta$.

From our point of view one of the most interesting features in type
II seesaw derives from the fact that the SU(2)$_L$ triplet $\Delta$
interacts with the SM Higgs doublet via both cubic and quartic
scalar couplings. These, as we will show in this letter, can have
far reachings implications for the SM bounds on $M_H$, which can be
studied by employing the coupled renormalization group
equations(RGEs) involving the SM Higgs doublet $\phi$ and $\Delta$.
We find, in particular, that for a plausible choice of parameters,
the SM Higgs boson mass $M_H$ can be as low as the LEP2 bound of
114.4 GeV.

This is in sharp contrast with type I seesaw in which the lower
 bound of 127 GeV for $M_H$ is increased by the presence of
 Dirac Yukawa coupling(s) involving right-handed neutrinos
 \cite{HMass-typeI}. With a Dirac Yukawa coupling equal, say,
 to unity, the vacuum stability bound is increased to 156 GeV,
 while the perturbativity bound remains close to 170 GeV.
The key difference is that in type II seesaw a mass
 for the SM Higgs boson, say in the range of 115 - 127 GeV,
 is easily realized, which is not the case for type I seesaw.

Before moving to the technical part let us note that $\Delta$
contains doubly charged particles which, if not too heavy, may be
produced at the LHC and Tevatron \cite{Triplet-LHC}. It is amusing
that with type II seesaw, a 'light' SM Higgs is consistent with
relatively light $\Delta$ particles with masses around a few hundred
GeV. Of course, it may turn out that the mass scale for $\Delta$
lies well above the TeV range, in which case we will only find the
'light' Higgs boson at the LHC.

We begin by introducing a triplet Higgs scalar $\Delta$,
 which transforms as $({\bf 3}, 1)$
 under the electroweak gauge group SU(2)$_L\times$U(1)$_Y$:
\bea
 \Delta=\frac{\sigma^i}{\sqrt{2}}
   \Delta_i=\left(
 \begin{array}{cc}
    \Delta^+/\sqrt{2} & \Delta^{++}\\
    \Delta^0 & -\Delta^+/\sqrt{2}\\
 \end{array}\right) .
\eea
 The scalar potential involving both the SM Higgs doublet
 and the triplet Higgs is given by
 (throughout this paper, we follow the notation
 of Ref.~\cite{Schmidt}, except that we employ
 lower case Greek letters for dimensionless couplings)
\bea
 V(\Delta, \phi) &=&
 -m_\phi^2 (\phi^\dagger \phi)
  + \frac{\lambda}{2} (\phi^\dagger \phi)^2 \nonumber\\
&&+ M_\Delta^2 \tr \left( \Delta^\dagger \Delta \right)
 + \frac{\lambda_1}{2} \left(\tr \Delta^\dagger \Delta \right)^2
 + \frac{\lambda_2}{2}\left[
  \left(\tr \Delta^\dagger \Delta \right)^2
 -\tr \left(\Delta^\dagger \Delta \Delta^\dagger \Delta \right)
 \right] \nonumber\\
&&+
 \lambda_4 \phi^\dagger \phi \; \tr \left(\Delta^\dagger\Delta\right)
 + \lambda_5 \phi^\dagger
 \left[\Delta^\dagger, \Delta\right] \phi
 + \left[ \frac{\Lambda_6}{\sqrt{2}}
  \phi^T i\sigma_2 \Delta^\dagger \phi +{\rm h.c.} \right],
 \label{H-D-Potential}
\eea where $\phi$ $({\bf 2}, 1/2)$ is the SM Higgs doublet. Without
loss of generality the coupling constants $\lambda_i$ are taken to
be real
 through a phase rotation of $\Delta$. Note that we define a
 dimensionless parameter $\lambda_6 \equiv \Lambda_6/M_\Delta$.
The triplet Higgs has a Yukawa coupling with the lepton doublets
 ($\ell_L^i$, with generation index $i$) of the form,
\bea
{\cal L}_\Delta =
 -\frac{1}{\sqrt{2}}\left(Y_\Delta\right)_{ij}
  \ell_L^{Ti} \mathrm{C} i \sigma_2 \Delta \ell_L^j +{\rm h.c.},
\label{Yukawa}
\eea
where $\mathrm{C}$ is the Dirac charge conjugate matrix
and $\left(Y_\Delta\right)_{ij}$ denotes elements
of the Yukawa matrix.

Assuming the hierarchy $M_Z \ll M_\Delta$ and
 integrating out the heavy triplet Higgs,
 we obtain a low energy effective potential for the SM doublet,
\bea
 V(\phi)_{\rm eff}  =
   -m_\phi^2 (\phi^\dagger \phi)
   + \frac{1}{2} \left( \lambda - \lambda_6^2 \right)
   (\phi^\dagger \phi)^2 .
\eea
Below $M_{\Delta}$ the SM Higgs quartic coupling is given by
\cite{matching} \bea
  \lambda_{\rm SM}= \lambda - \lambda_6^2 .
 \label{matching}
\eea
For a given $\lambda_6$, the Higgs quartic coupling is shifted
 down to the SM coupling by $\lambda_6^2$
 through this matching condition at $\mu = M_\Delta$,
 so that the resulting Higgs boson mass, as we will show, is lowered.

A non-zero VEV($v=246.2$ GeV) for the Higgs doublet induces
 a tadpole term for $\Delta$ via the last term in
Eq.~(\ref{H-D-Potential}). A non-zero VEV of the triplet Higgs is
thereby generated,
 $\langle \Delta \rangle \sim \lambda_6 v^2/M_\Delta$,
 which then provides the desired neutrino masses using Eq.~(\ref{Yukawa}).

Note that the triplet Higgs VEV contributes
 to the weak boson masses
 and alters the $\rho$-parameter from the SM prediction,
 $\rho \approx 1$, at tree level.
The current precision measurement \cite{PDG}
 constrains this deviation to be in the range,
 $ \Delta \rho =\rho -1 \simeq
 \langle \Delta \rangle/v  \lesssim 0.01$,
 so that $\lambda_6 \lesssim 0.01 M_\Delta/v$.
This constraint is especially relevant when we consider
 $M_\Delta\sim $ TeV, in which case
 the region $\lambda_6 \gtrsim 0.1$ is excluded.

We are now ready to analyze the Higgs boson mass bounds
 from vacuum stability and perturbativity constraints
 in the presence of type II seesaw mechanism.
There are several new parameters
 $\lambda_1$, $\lambda_2$, $\lambda_4$, $\lambda_5$,
 $\lambda_6$, and $Y_\Delta$,
 which potentially affect the RGE running of the Higgs quartic coupling
 and thus the corresponding Higgs boson mass. We have already noted
 the important role that $\lambda_6$ plays via the matching
 condition in Eq. (5). It turns out that both $\lambda_4$ and
 $\lambda_5$ will also play an important role through their contributions to
 the renormalization group evolution of the Higgs quartic coupling.
 They help to lower both the vacuum  stability and perturbativity
 bounds on the SM Higgs mass.
In our analysis, we employ two-loop RGEs
 for the SM couplings and one-loop RGEs
 for the new couplings associated with the type II seesaw scenario.

For renormalization scale $ \mu < M_\Delta $, the triplet Higgs is
decoupled. For the three SM gauge couplings, we have
\bea
 \frac{d g_i}{d \ln \mu} =
 \frac{b_i}{16 \pi^2} g_i^3 +\frac{g_i^3}{(16\pi^2)^2}
\sum_{j=1}^3b_{ij}g_j^2, \label{gauge} \eea
 where
$g_i$ ($i=1,2,3$) are the SM  gauge couplings  and
\bea
b_i = \left(\frac{41}{10},-\frac{19}{6},-7\right),~~~~~~~~
 { b_{ij}} =
 \left(
  \begin{array}{ccc}
  \frac{199}{50}& \frac{27}{10}&\frac{44}{5}\\
 \frac{9}{10} & \frac{35}{6}&12 \\
 \frac{11}{10}&\frac{9}{2}&-26
  \end{array}
 \right).
\label{beta}
\eea
The top quark pole mass is taken to be
 the central value $M_t= 170.9$ GeV, \cite{Tevatron},
with
 $(\alpha_1, \alpha_2, \alpha_3)=(0.01681, 0.03354, 0.1176)$
 at the Z-pole ($M_Z$) \cite{PDG}.
For the top Yukawa coupling $y_t$, we have \cite{RGE},
\bea \label{ty}
 \frac{d y_t}{d \ln \mu}
 = y_t  \left(
 \frac{1}{16 \pi^2} \beta_t^{(1)} + \frac{1}{(16 \pi^2)^2} \beta_t^{(2)}
 \right).
\eea
Here the one-loop contribution is
\bea
 \beta_t^{(1)} =  \frac{9}{2} y_t^2 -
  \left(
    \frac{17}{20} g_1^2 + \frac{9}{4} g_2^2 + 8 g_3^2
  \right) ,
\label{topYukawa-1}
\eea
while the two-loop contribution is given by
\bea
\beta_t^{(2)} &=&
 -12 y_t^4 +   \left(
    \frac{393}{80} g_1^2 + \frac{225}{16} g_2^2  + 36 g_3^2
   \right)  y_t^2  \nonumber \\
 &&+ \frac{1187}{600} g_1^4 - \frac{9}{20} g_1^2 g_2^2 +
  \frac{19}{15} g_1^2 g_3^2
  - \frac{23}{4}  g_2^4  + 9  g_2^2 g_3^2  - 108 g_3^4 \nonumber \\
 &&+ \frac{3}{2} \lambda^2 - 6 \lambda y_t^2 .
\label{topYukawa-2}
\eea
In solving Eq.~(\ref{ty}),
 the initial top Yukawa coupling at $\mu=M_t$
 is determined from the relation
 between the pole mass and the running Yukawa coupling
 \cite{Pole-MSbar}, \cite{Pole-MSbar2},
\bea
  M_t \simeq m_t(M_t)
 \left( 1 + \frac{4}{3} \frac{\alpha_3(M_t)}{\pi}
          + 11  \left( \frac{\alpha_3(M_t)}{\pi} \right)^2
          - \left( \frac{m_t(M_t)}{2 \pi v}  \right)^2
 \right),
\eea
 with $ y_t(M_t) = \sqrt{2} m_t(M_t)/v$, where $v=246.2$ GeV.
Here, the second and third terms in the parenthesis correspond to
 one- and two-loop QCD corrections, respectively,
 while the fourth term comes from the electroweak corrections at one-loop level.
The numerical values of the third and fourth terms
 are comparable (signs are opposite).
The electroweak corrections at two-loop level and
 the three-loop QCD corrections \cite{Pole-MSbar2},
 are of comparable and sufficiently small magnitude \cite{Pole-MSbar2}
 to be safely ignored.

The RGE  for the Higgs quartic coupling is given by \cite{RGE},
\bea
\frac{d \lambda}{d \ln \mu}
 =   \frac{1}{16 \pi^2} \beta_\lambda^{(1)}
   + \frac{1}{(16 \pi^2)^2}  \beta_\lambda^{(2)},
\label{self}
\eea
with
 \bea
 \beta_\lambda^{(1)} &=& 12 \lambda^2 -
 \left(  \frac{9}{5} g_1^2+9 g_2^2  \right) \lambda
 + \frac{9}{4}
 \left(
 \frac{3}{25} g_1^4 + \frac{2}{5} g_1^2 g_2^2 +g_2^4
 \right) \nonumber \\
&&+ 12 y_t^2 \lambda  - 12 y_t^4 ,
\label{self-1}
\eea
and
\bea
  \beta_\lambda^{(2)} &=&
 -78 \lambda^3  + 18 \left( \frac{3}{5} g_1^2 + 3 g_2^2 \right) \lambda^2
 - \left( \frac{73}{8} g_2^4  - \frac{117}{20} g_1^2 g_2^2
 + \frac{2661}{100} g_1^4  \right) \lambda - 3 \lambda y_t^4
 \nonumber \\
 &&+ \frac{305}{8} g_2^6 - \frac{289}{40} g_1^2 g_2^4
 - \frac{1677}{200} g_1^4 g_2^2 - \frac{3411}{1000} g_1^6
 - 64 g_3^2 y_t^4 - \frac{16}{5} g_1^2 y_t^4
 - \frac{9}{2} g_2^4 y_t^2
 \nonumber \\
 && + 10 \lambda \left(
  \frac{17}{20} g_1^2 + \frac{9}{4} g_2^2 + 8 g_3^2 \right) y_t^2
 -\frac{3}{5} g_1^2 \left(\frac{57}{10} g_1^2 - 21 g_2^2 \right)
  y_t^2  - 72 \lambda^2 y_t^2  + 60 y_t^6.
\label{self-2}
\eea
The Higgs boson pole mass $M_H$ is determined
 by the relation to the running Higgs quartic coupling
through the one-loop matching condition \cite{HiggsPole} , \bea
 \lambda(M_H) \; v^2 = M_H^2 \left( 1+ \Delta_h(M_H) \right),
\eea
where
\bea
 \Delta_h(M_H) = \frac{G_F}{\sqrt{2}} \frac{M_Z^2}{16 \pi^2}
 \left[
   \frac{M_H^2}{M_Z^2} f_1\left(\frac{M_H^2}{M_Z^2}\right)
 + f_0\left(\frac{M_H^2}{M_Z^2}\right)
 + \frac{M_Z^2}{M_H^2} f_{-1}\left(\frac{M_H^2}{M_Z^2}\right)
 \right]
\eea
The functions are given by
\bea
f_1(\xi) &=&
   \frac{3}{2}  \ln \xi - \frac{1}{2} Z\left(\frac{1}{\xi}\right)
 -Z\left(\frac{c_w^2}{\xi}\right) -\ln c_w^2
  +\frac{9}{2} \left( \frac{25}{9} - \frac{\pi}{\sqrt{3}}
  \right), \nonumber\\
f_0(\xi)  &=& - 6\ln\frac{M_H^2}{M_Z^2}
  \left[ 1 +2 c_w^2 -2 \frac{M_t^2}{M_Z^2} \right]
 +\frac{3 c_w^2 \xi}{\xi-c_w^2} \ln \frac{\xi}{c_w^2}
  +2 Z\!\left( \frac{1}{\xi} \right) \nonumber\\
 &&+ 4 c_w^2 Z\left( \frac{c_w^2}{\xi} \right)
   +\left(\frac{3 c_w^2}{s_w^2} +12 c_w^2\right) \ln c_w^2
  -\frac{15}{2} \left( 1 +2 c_w^2 \right) \nonumber\\
 &&- 3\frac{M_t^2}{M_Z^2} \left[
     2 Z\left( \frac{M_t^2}{M_Z^2 \xi} \right)
    +4 \ln \frac{M_t^2}{M_Z^2} -5 \right], \nonumber\\
f_{-1}(\xi) &=& 6 \ln \frac{M_H^2}{M_Z^2}
  \left[ 1 +2 c_w^4 -4\frac{M_t^4}{M_Z^4} \right]
 -6 Z\left( \frac{1}{\xi} \right)-12 c_w^4 Z\left( \frac{c_w^2}{\xi} \right)
 -12 c_w^4 \ln c_w^2 \nonumber\\
 && + 8\left( 1 +2 c_w^4 \right) +24 \frac{M_t^4}{M_Z^4}
  \left[\ln \frac{M_t^2}{M_Z^2} -2 + Z\left( \frac{M_t^2}{M_Z^2 \xi} \right)
\right],
\eea
with $s_w^2 = \sin^2 \theta_W$, $ c_w^2 = \cos^2 \theta_W$
 ($\theta_W$ denotes the weak mixing angle) and
\bea
 Z(z) = \left\{
  \begin{array}{cc}
    2 A \arctan(1/A) & (z > 1/4 ) \\
    A \ln \left[ (1+A)/(1-A) \right] & (z < 1/4 ) ,
 \end{array}\right.
\eea
 with $ A = \sqrt{ \left| 1 - 4 z \right| }$.

For $ \mu \geq M_\Delta$, the triplet Higgs contributes
 to the one-loop RGEs.
Consequently,  we replace $b_i$ in Eq.~(\ref{beta}) with
\bea
  b_i = \left(\frac{47}{10},-\frac{5}{2},-7\right).
\eea
The RGE for the top Yukawa coupling is unchanged.

The RGEs for the new couplings
 in Eqs.~(\ref{H-D-Potential}) and (\ref{Yukawa})
 have been calculated in Ref.~\cite{Chao-Zhang}
 and more recently in Ref.~\cite{Schmidt}.
The RGE of the Higgs quartic coupling acquires
 a new entry in Eq.(\ref{self-1}),
\bea
 \beta_\lambda^{(1)} \to
 \beta_\lambda^{(1)}+ 6 \lambda_4^2 + 4 \lambda_5^2.
\label{self-1New} \eea

Note that in Eq. (20) the contributions from the couplings
$\lambda_4$ and $\lambda_5$ are both positive. This feature will be
crucial in lowering both the vacuum stability and perturbativity
bounds with type II seesaw.

For the remaining couplings we have
\cite{Schmidt}, \bea \label{Lam1}
 16\pi^2 \frac{d \lambda_1}{d \ln \mu} &=&
  -\left( \frac{36}{5} g_1^2 + 24g_2^2 \right) \lambda_1
  +\frac{108}{25}g_1^4 +18g_2^4 + \frac{72}{5}g_1^2g_2^2 \nonumber\\
  && + 14\lambda_1^2 +4 \lambda_1 \lambda_2 + 2 \lambda_2^2
  + 4 \lambda_4^2+4\lambda_5^2
  + 4 \tr\left[{\bf S}_\Delta \right] \lambda_1
  - 8 \tr\left[{\bf S}_\Delta^2 \right],  \\
\label{Lam2}
 16\pi^2 \frac{d \lambda_2}{d \ln \mu} &=&
  - \left( \frac{36}{5}g_1^2 + 24g_2^2 \right)\lambda_2
  +12g_2^4 -\frac{144}{5}g_1^2g_2^2 \nonumber\\
  && + 3 \lambda_2^2 +12 \lambda_1 \lambda_2 - 8 \lambda_5^2
  + 4 \tr \left[{\bf S}_\Delta  \right]\lambda_2
  + 8 \tr\left[{\bf S}_\Delta^2 \right],  \\
\label{Lam3}
 16\pi^2\ \frac{d \lambda_4}{d \ln \mu} &=&
 - \left( \frac{9}{2}g_1^2 + \frac{33}{2}g_2^2 \right)\lambda_4
 +\frac{27}{25}g_1^4 +6g_2^4 \nonumber\\
 && + \left( 8 \lambda_1 + 2 \lambda_2 + 6 \lambda+4\lambda_4 +6 y_t^2
 + 2\tr\left[{\bf S}_\Delta \right] \right) \lambda_4
 +8 \lambda_5^2-4\tr\left[ {\bf S}_\Delta^2 \right], \\
\label{Lam4}
 16\pi^2 \frac{d \lambda_5}{d \ln \mu} &=&
 -\frac{9}{2}g_1^2\lambda_5-\frac{33}{2}g_2^2\lambda_5
 -\frac{18}{5}g_1^2g_2^2 \nonumber\\
 &&+ \left(  2\lambda_1-2\lambda_2+2 \lambda + 8 \lambda_4 + 6 y_t^2
 + 2 \tr\left[{\bf S}_\Delta \right] \right) \lambda_5
 + 4\tr\left[{\bf S}_\Delta^2 \right] .
\eea
Here, ${\bf S}_\Delta=Y_\Delta^\dagger Y_\Delta $
 and its corresponding RGE is given by
\bea
 16\pi^2 \frac{d {\bf S}_\Delta}{d \ln \mu} =
 6 \; {\bf S}_\Delta^2
 -3 \left( \frac{3}{5} g_1^2 +3 g_2^2 \right) {\bf S}_\Delta
 + 2 \tr[{\bf S}_\Delta] {\bf S}_\Delta .
\label{S_Delta}
\eea

In our analysis, we do not consider the RGE for $\lambda_6$
 because, at one-loop level, it is decoupled from the other RGEs.
We assume though that like all other couplings,
 it remains in the perturbative region throughout.
However, note that $\lambda_6$ plays an important role
 in our analysis through the matching condition
 for the Higgs quartic coupling between high and low energies.
The SM Higgs quartic coupling as the low energy effective coupling
 is defined in Eq.~(\ref{matching}).
Therefore, the RGE solution of the SM Higgs quartic coupling
 is connected to the solution for the Higgs quartic coupling
 at high energies through the matching condition
 at $\mu = M_\Delta$ \cite{matching}.
For a given $\lambda_6$, the Higgs quartic coupling is shifted
 down to the SM coupling by $\lambda_6^2$
 through this matching condition at $\mu = M_\Delta$,
 so that the resulting Higgs boson mass is lowered.

Next we analyze the RGEs numerically and show how much
 the vacuum stability and perturbativity bounds on Higgs boson mass
 are altered in the presence of type II seesaw.
In addition to the matching condition,
 $\lambda_{4,5}$ have important effects on
 the running of the Higgs quartic coupling
 since they appear in Eq.~(\ref{self-1New}).
Their contributions are positive and these couplings work
 so as to reduce the Higgs quartic coupling at low energies.
Consequently, we can expect that the resultant Higgs boson mass
 tends to be reduced from the effects of $\lambda_6$
 and $\lambda_{4,5}$ in type II seesaw.
Although the other parameters $\lambda_{1,2}$ and $Y_\Delta$
 are not involved in the RGE of Higgs quartic coupling,
 they, of course, do affect the Higgs boson mass
 through their RGEs coupled with $\lambda_{4,5}$.
 To keep the discussion as simple as possible we proceed as follows.
We first investigate the Higgs boson mass bounds by varying
 $\lambda_6$, while keeping the other(non-SM) parameters fixed.
We show that a Higgs mass as low as the LEP 2 bound can be
 realized in this case.
With the seesaw scale at TeV, however, the lower bound
 on the Higgs mass is closer to 120 GeV.
For our next examples we vary $\lambda_5$, keeping
 the other(non-SM) parameters fixed.
This case yields a lower bound of 114.4 GeV for the Higgs boson
 mass which is compatible with a low (~ TeV or so) seesaw scale.

Fixing the cutoff scale as $M_{Pl}=1.2 \times 10^{19}$ GeV,
 we define the vacuum stability bound as the lowest Higgs mass
 given by the running Higgs quartic coupling which satisfies
 the condition $\lambda(\mu) \geq 0$ for any scale
 between $M_H \leq \mu \leq M_{Pl}$.
On the other hand, the perturbativity bound is defined as
 the highest Higgs boson mass given by
 the running Higgs quartic coupling which satisfies the condition
 $\lambda(\mu) \leq \sqrt{4 \pi}$ for any scale
 between $M_H \leq \mu \leq M_{Pl}$.

In Fig.~1, the running Higgs mass defined as $\sqrt{\lambda(\mu)} v$
 for the vacuum stability bound is depicted
 for various $\lambda_6$ and a fixed seesaw scale $M_\Delta=10^{12}$ GeV.
Here we took simple inputs as
 $\lambda_1=\sqrt{4 \pi}$, $\lambda_2=-1$, $\lambda_4=\lambda_5=0$
 and $Y_\Delta=0$.
Fig.~2 shows a running Higgs mass
 for the perturbativity bound for various $\lambda_6$
 with the same inputs for the other parameters as in Fig.~1.
We find that as $\lambda_6$ is increased, the vacuum  stability and
 perturbativity bounds eventually merge.
The corresponding Higgs mass coincides, it turns out,
 with the vacuum stability bound for the SM Higgs boson mass
 obtained with a cutoff scale $\Lambda=M_\Delta$.

In other words, the window for the Higgs boson mass
 between the vacuum stability and perturbative bounds
 becomes narrow and is eventually closed
 as $\lambda_6$ becomes sufficiently large.
Fig.~3 shows the window for the Higgs boson pole mass
 versus $\lambda_6$ for various $M_\Delta$. For a suitable choice
 of $\lambda_6$ and $M_\Delta$, values for $M_H$ close to the LEP2
 bound are easily realized. Note that with $M_\Delta$ of order a TeV
 or so, the lower bound on the Higgs mass is close to 120 GeV,
 which is still well below the standard value of 127 GeV
 in the absence of type II  seesaw.

In Fig.~4, a running Higgs mass defined as $\sqrt{\lambda(\mu)} v$
 for the vacuum stability bound is depicted
 for various $\lambda_5$ and
 a fixed seesaw scale $M_\Delta=10^{12}$ GeV.
Here we took sample inputs as
 $\lambda_1=\sqrt{4 \pi}$, $\lambda_2=-1$, $\lambda_4=0$,
 $\Lambda_6=0$ and $Y_\Delta=0$.
Fig.~5 shows a running Higgs mass
 for the perturbativity bound for various $\lambda_5$
 with the same inputs for the other parameters as in Fig.~4.
This time we find that as $\lambda_5$ is raised,
 the vacuum stability and perturbativity bounds eventually merge.
The corresponding Higgs mass coincides with the vacuum stability
bound
 for the SM Higgs boson with a cutoff scale $\Lambda=M_\Delta$.
Namely, the window for the Higgs boson mass
 between the vacuum stability and perturbative bounds
 becomes narrow and is eventually closed
 when $\lambda_5$ becomes sufficiently large.
Fig.~6 shows the window for the Higgs boson pole mass
 versus $\lambda_5$ for various $M_\Delta$.
Note that with $M_{\Delta}$ = 1 TeV and $\lambda_5$ close to 0.1,
the vacuum  stability bound for $M_H$ coincides with the LEP2 bound.

Some comments regarding our analysis are in order here. We employ a
set of input parameters such that all couplings remain in the
perturbative regime for
 $M_\Delta \leq \mu \leq M_{Pl}$. For
example, for an input parameter set, it may happen
 that $\lambda_{1,2}$ exceed the perturbative regime
 at a scale $M_\Delta \leq \mu \leq M_{Pl}$.
We checked that the sample parameter set used in Fig.~1-6
 did not cause such theoretical inconsistency.
To reproduce the current neutrino oscillation data
 through the type II seesaw mechanism,
 $\lambda_6 Y_\Delta$ should not be zero.
A tiny $\lambda_6$ would require
 a very large $Y_\Delta$, which may cause theoretical inconsistency
 for the RGEs of $\Lambda_i$ (see Eq.~(\ref{Lam1})-(\ref{Lam4})).
For this reason, we do not strictly
 set $\lambda_6=0$ and $Y_\Delta=0$ in Figs.~1-6.
However, we can check that the impact of $Y_\Delta$ in Fig.~3
 is negligible when $Y_\Delta \leq 0.1$.
In Fig.~6, the effects from $Y_\Delta$ and $\lambda_6$
 are negligible with $Y_\Delta,\; \lambda_6 \leq 0.1$.

In conclusion, we have considered the potential impact of
 type II seesaw on the vacuum stability and perturbativity bounds
 on the Higgs boson mass of the SM. There are two main effects that
we have considered. One is the tree level matching condition for the
SM Higgs quartic coupling induced by the coupling $\Lambda_6$ at the
seesaw (triplet mass) scale. The implications for the SM Higgs mass
are then studied using the appropriate RGEs. In our second set of
examples a different coupling, namely $\lambda_5$, plays an
essential role, with $\Lambda_6$ essentially subdominant.
In both cases we have identified regions of the parameter space
for which the lower bound on the SM Higgs boson mass lies well
below 127 GeV, the value obtained in the absence of type II seesaw.
Perhaps the most interesting result from our analysis
 is that with type II seesaw the SM Higgs boson can have a mass
 as low as the LEP 2 bound of 114.4 GeV.
This can be achieved for plausible values of the parameters
 and with the mass scale for the triplet in the TeV range.
Thus, it is exciting to speculate that the SM Higgs boson,
 when found at the LHC, may be accompanied by additional scalar
 fields with some carrying two units of electric charge.

\section*{Acknowledgments}
This work is supported in part by
 the DOE Grant \# DE-FG02-91ER40626 (I.G. and Q.S.),
 and the Grant-in-Aid for Scientific Research from the Ministry
 of Education, Science and Culture of Japan,
 \#18740170 (N.O.).


\newpage
\begin{figure}[t]
\includegraphics[scale=1.2]{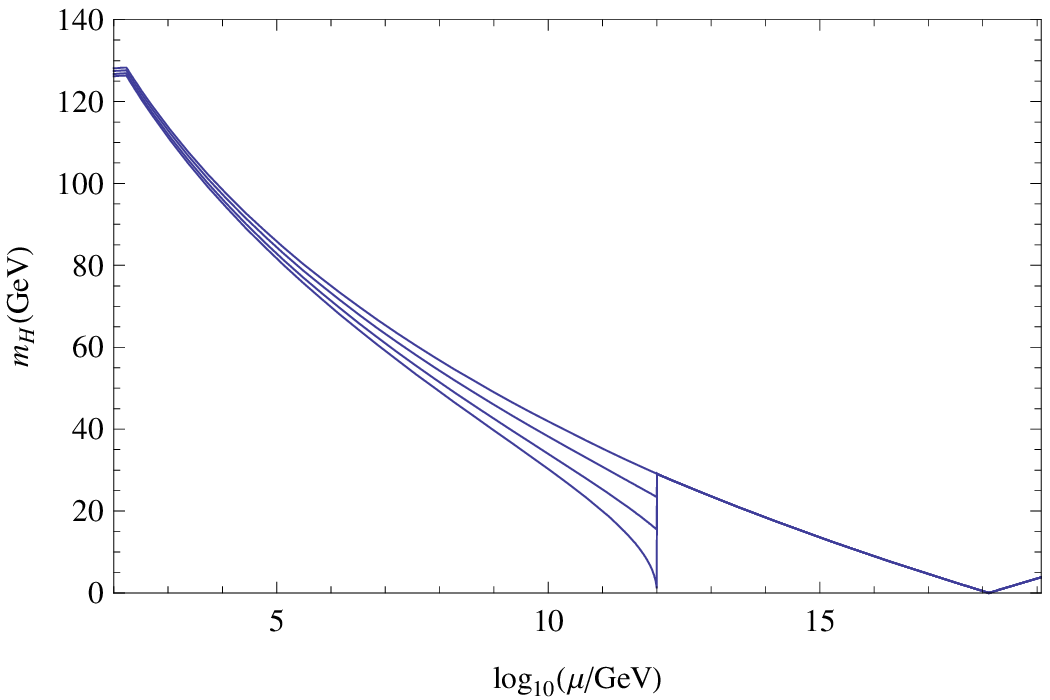}
\caption{Evolution of running Higgs boson mass $(m_h(\mu=
{\sqrt{\lambda(\mu)} v}))$ corresponding to the vacuum stability
bound
 for various $\lambda_6$ and $M_\Delta=10^{12}$ GeV .
Each line corresponds to
 $\lambda_6=0$, $0.07$, $0.1$ and $0.118$,
 from top to bottom.
}
\end{figure}
\begin{figure}[t]
\includegraphics[scale=1.2]{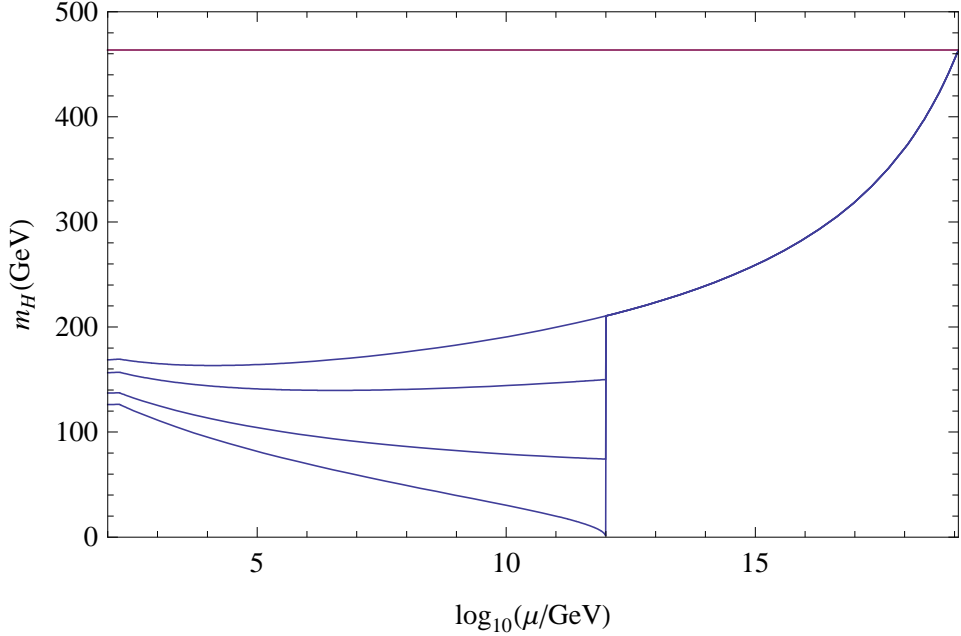}
\caption{ Evolution of running Higgs boson mass $(m_h(\mu=
{\sqrt{\lambda(\mu)} v}))$  corresponding  to the perturbativity
bound
 for various $\lambda_6$ and $M_\Delta=10^{12}$ GeV.
Each line corresponds to
 $\lambda_6=0$, $0.6$, $0.8$ and $0.855$
 from top to bottom.
The horizontal line corresponds to
 $M_H=(4 \pi)^{1/4} v= 464$ GeV.
}
\end{figure}
\begin{figure}[t]
\includegraphics[scale=1.2]{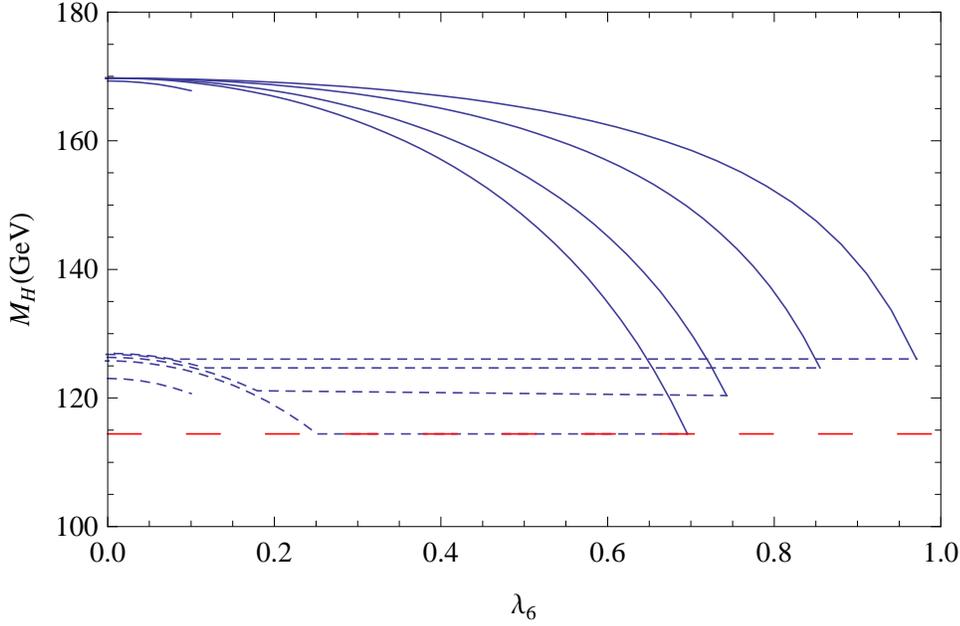}
\caption{ The perturbativity (solid) and vacuum stability (dotted)
 bounds on the Higgs boson pole mass $M_H$ versus $\lambda_6$ for various $M_\Delta$.
Each solid and dashed lines correspond to
 $M_\Delta=10^{14}$, $10^{12}$, $10^9$,
 $1.14 \times 10^7$ and $10^3$ GeV, from top to bottom.
The results for $M_\Delta=1$ TeV are shown
 only in the region $\lambda_6 \leq 0.1$
 consistent with the $\rho$-parameter measurement.
The dashed horizontal line denotes the LEP2 bound $M_H=114.4$ GeV.
}
\end{figure}
\begin{figure}[t]
\includegraphics[scale=1.2]{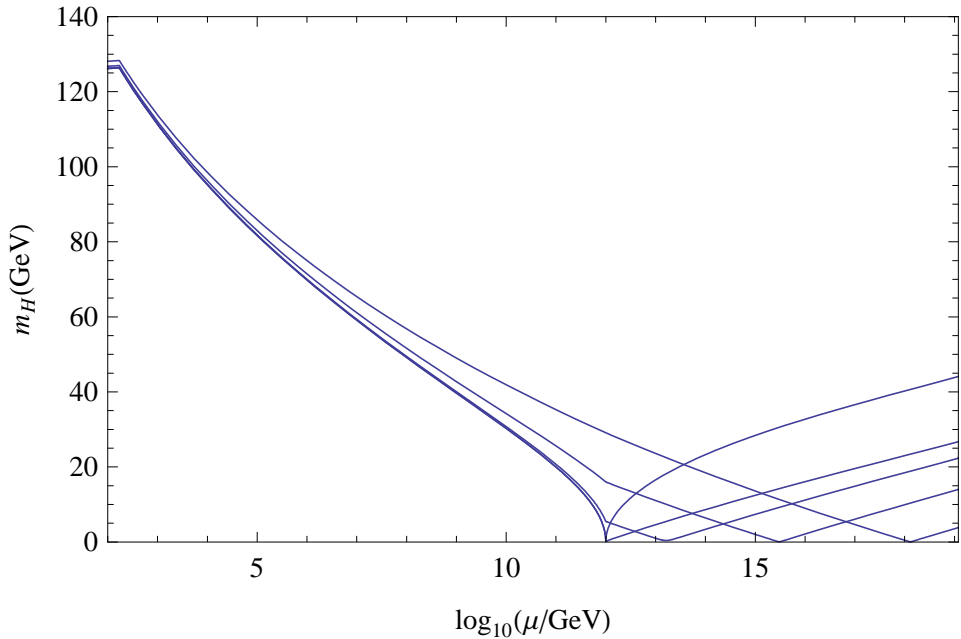}
\caption{ Evolution of running Higgs boson mass corresponding  to
the vacuum stability  bound for various value $\lambda_5$ and
$M_\Delta=10^{12}$ GeV. Each line corresponds to
 $\lambda_5=0.3$, $0.22$, $0.2$, $0.15$ and $0$,
 from top to bottom at $\log_{10}(\mu/{\rm GeV})=19$.
}
\end{figure}
\begin{figure}[t]
\includegraphics[scale=1.2]{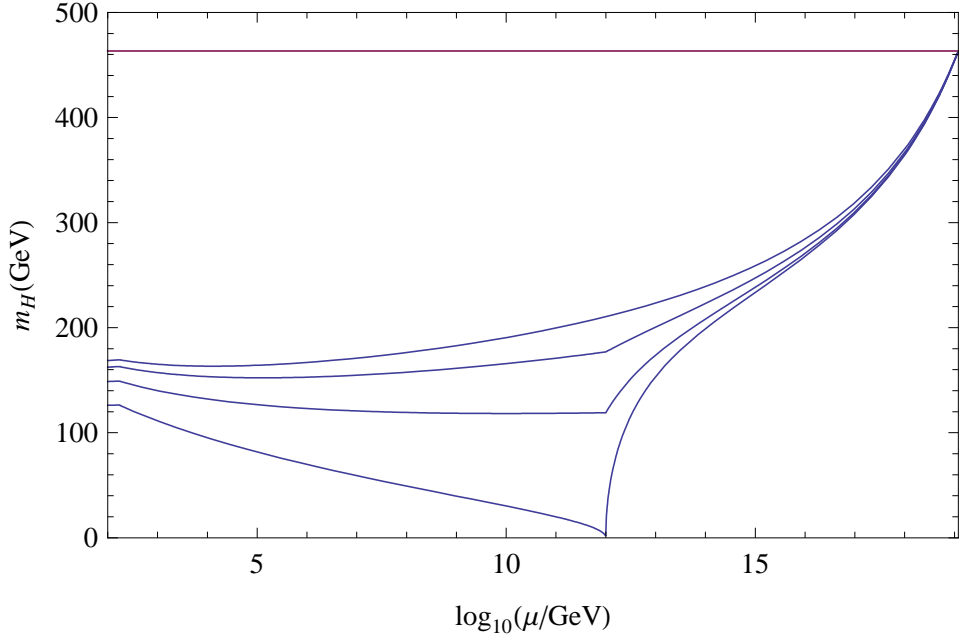}
\caption{ Evolution of running  Higgs boson mass corresponding  to
the perturbativity bound
 for various $\lambda_5$ and $M_\Delta=10^{12}$ GeV.
Each line corresponds to
 $\lambda_5=0$, $1.0$, $1.25$ and $1.35$, from top to bottom.
The horizontal line corresponds to
 $M_H=(4 \pi)^{1/4} v= 464$ GeV.
}
\end{figure}
\begin{figure}[t]
\includegraphics[scale=1.2]{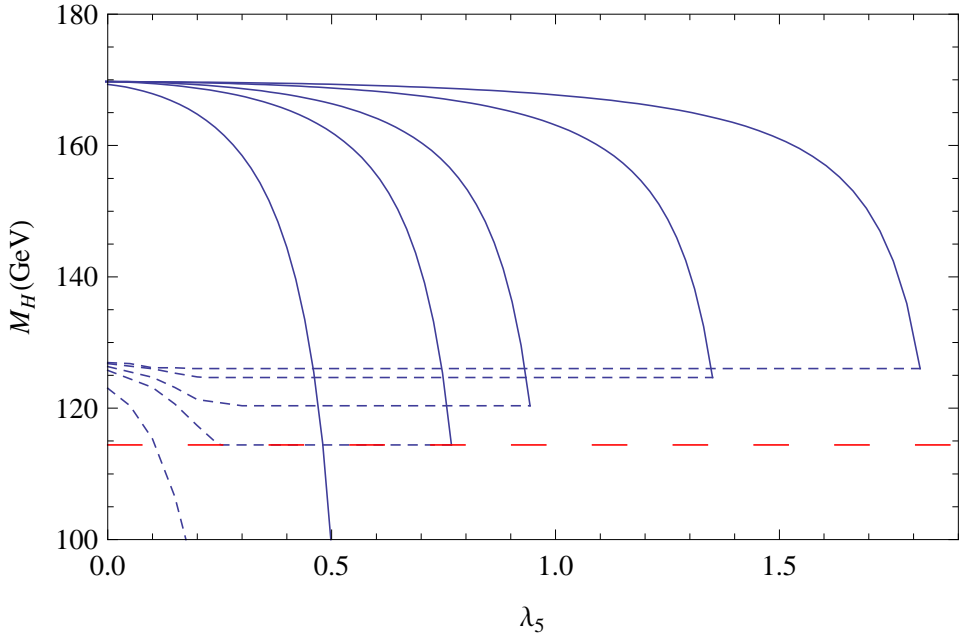}
\caption{ The perturbativity (solid) and vacuum stability (dotted)
 bounds versus $\lambda_5$ for various $M_\Delta$.
Each solid and dashed lines correspond to
 $M_\Delta=10^{14}$, $10^{12}$, $10^9$,
 $1.14 \times 10^7$ and $10^3$ GeV, from top to bottom.
The dashed horizontal  line denotes the LEP2 bound $M_H=114.4$ GeV.
}
\end{figure}

\end{document}